\documentstyle[referee,epsf]{mn} 

\newcommand\etal{et al. }
\newcommand\refer{\par \noindent\hangindent=3pc \hangafter=1}

\title[ROSAT PSPC spectra of X--ray selected Narrow Emission Line
Galaxies]{ROSAT PSPC spectra of X--ray selected Narrow Emission Line Galaxies}

\author[Romero Colmenero \etal] {E. Romero-Colmenero\(^{1}\),
G. Branduardi-Raymont\(^{1}\), F. J. Carrera\(^{1}\), \and L. R. Jones\(^{2}\),
K. O. Mason\(^{1}\), I. M.  McHardy\(^{3}\), J. P. D. Mittaz\(^{1}\)
\\ 
\(^{1}\) Mullard Space Science Laboratory, University
College London, Holmbury St Mary, Dorking, Surrey, UK \\ 
\(^{2}\) Code 660.2, NASA/Goddard Space Flight Center, Greenbelt, Maryland, 
USA\\
\(^{3}\) Physics Department, University of Southampton, Highfield, Southampton,
UK} 

\date{}
\begin{document}

\maketitle

\begin{abstract}

\noindent 

We analyse the ROSAT PSPC spectrum  of 19 X--ray selected Narrow Emission
Line Galaxies (NELGs) discovered during the optical identification of  sources
in the ROSAT UK Deep Survey. Their properties are compared to those of broad
line Active Galactic Nuclei (AGN) in the same sample.

Counts in three spectral bands have been extracted for all the sources, and
have been fitted with a power-law model assuming the Galactic value for
$N_{H}$. The average slope of NELGs is $\alpha$= 0.45 $\pm$ 0.09 , whilst for
the AGN it is $\alpha $= 0.96 $\pm$ 0.03. The power-law model is a good fit for
$\sim$ 90\% of NELGs and $\sim$ 75\% of AGN. 

Recent work shows that the fractional surface density of NELGs increases with
respect to AGN at faint fluxes. Thus they are expected to be an important
component of the residual soft ($< 2$ keV) X--ray background.  The slope of
the X--ray background ($\alpha \sim $ 0.4, 1-10 keV) is harder than that of AGN
($\alpha \sim$ 1) but our results show that it is consistent with the summed
spectrum of the NELGs  in the deep survey ($\alpha \sim$ 0.4). This may finally
reconcile  the spectrum of the background with the properties of the sources
that constitute it.

\end{abstract}

\section{Introduction}

The soft ($< 2$ keV) X--ray background is thought to arise from the integrated
signal of individual unresolved sources (Fabian and Barcons, 1992, and
references therein). Although broad line AGN are known to be the main
contributors to the soft X--ray background at higher fluxes (Shanks \etal 1991;
Boyle \etal 1994), their spectral shape (energy index,
$\alpha \sim 1$, Maccacaro \etal 1988) seems too steep to match that of the 
background ($\alpha \sim$ 0.4, 1-10~keV, Gendreau \etal 1995, Chen \etal 1996).
Hence a new population of harder and fainter sources, with a steep Log N-Log S
and which do not contribute significantly above fluxes $\sim$ 10$^{-14}$ erg
cm$^{\rm -2}$ s$^{\rm -1}$, has been sought in order to resolve this spectral
paradox (Hasinger \etal 1993). Narrow Emission Line Galaxies (NELGs) seem to be
an attractive candidate for this new contributor to the X--ray background at
fainter fluxes ( Jones \etal 1995a, Boyle \etal 1995).

NELGs are generally defined as galaxies which possess only emission lines with
FWHM $< 1000$ km s$^{\rm -1}$ in their optical spectra. They can be
substantially brighter than normal galaxies in X--rays. Thus for example
Fabbiano (1989) finds X-ray luminosities in the range $\sim$ 10$^{38}$ -
10$^{42}$ erg s$^{-1}$ (0.2 - 3.5 keV) for normal galaxies, while our sample of
NELGs yields $\sim$ 10$^{41}$ - 2 x 10$^{43}$ erg s$^{-1}$ when extrapolated to
the same energy band.  The number ratio of NELGs has recently been found to
increase at faint X-ray fluxes (Jones \etal 1995a, Boyle \etal 1995), 
leading to the suggestion that they may be
important contributors to the unresolved fraction of the X--ray
background.  However, in order to solve the soft X--ray spectral paradox, NELGs
would need to have a flatter spectrum than the X--ray background, to compensate
for the steeper slope of AGN.

In this paper we examine X--ray data on faint NELGs to determine if their
integrated spectrum can reproduce the spectral shape of the X--ray background.
We present a total sample of 19 NELG from the UK Deep Survey, and compare their
spectra to that of broad line AGN from the same survey, and to the spectrum of 
the soft X--ray background. In section 2, the deep survey sample is described.
In section 3 we explain the data reduction and analysis process. We report and
discuss our results in section 4, and in section 5 we present our conclusions.
A preliminary report of this work was given in Romero-Colmenero \etal (1996).

\footnotesep=-1.5in
\begin{table*}
\caption {Spectral results on Deep Survey NELGs}
\begin{minipage}[t]{10in}   
\begin{tabular}{c c c c c c r @{\hspace{1mm}}c @{\hspace{1mm}} l}\hline

 ID & extraction & z & $\alpha$ & Hard Counts \footnote[1]{Corrected for PSF and
vignetting. The error is the uncertainty in the counts.

\noindent Note: All errors quoted in the table are 1 $\sigma$ deviation.} & Fx (0.5-2 keV) & 
 \multicolumn{3}{c}{Lx (0.5-2 keV)} \\ 
Number & radius (arcsec)& & & PHA channels 52-201 
& 10$^{\rm-14}$ erg cm$^{\rm -2}$ s$^{\rm -1}$ 
& \multicolumn{3}{c}{10$^{\rm40}$ erg s$^{\rm -1}$} \\  \hline
032 & 43 & 0.068 & -2.41 $\pm$ 0.40 & 101.98 $\pm$  13.03 & 
1.96 $\pm$ 0.43 &   34& $\pm$ &  7 \\
036 & 41 & 0.235 &  0.18 $\pm$ 0.26 &  89.68 $\pm$  11.49 & 
1.31 $\pm$ 0.20 &  328& $\pm$ & 49 \\
042 & 18 & 0.366 & -0.15 $\pm$ 0.88 &  50.71 $\pm$   9.78 & 
0.76 $\pm$ 0.22 &  430& $\pm$ &116 \\
043 & 25 & 0.382 &  0.81 $\pm$ 0.25 &  69.76 $\pm$  13.32 & 
0.99 $\pm$ 0.20 &  827& $\pm$ &167 \\ 
047 & 18 & 0.364 &  0.49 $\pm$ 0.56 &  36.48 $\pm$   8.79 & 
0.52 $\pm$ 0.15 &  357& $\pm$ &101 \\ 
051 & 54 & 0.062 &  0.78 $\pm$ 0.19 &  64.77 $\pm$  10.56 & 
0.92 $\pm$ 0.16 &   16& $\pm$ &  3 \\
060 & 26 & 0.580 &  0.51 $\pm$ 0.37 &  38.47 $\pm$   7.60 & 
0.55 $\pm$ 0.12 & 1062& $\pm$ &237 \\
067 & 54 & 0.554 & -1.85 $\pm$ 1.77 &  44.71 $\pm$   9.50 & 
0.81 $\pm$ 0.38 &  497& $\pm$ &234 \\ 
085 & 48 & 0.304 & -0.29 $\pm$ 1.35 &  19.89 $\pm$   7.50 & 
0.30 $\pm$ 0.17 &  113& $\pm$ & 64 \\
093 & 49 & 0.590 &  0.57 $\pm$ 0.47 &  36.88 $\pm$   8.81 & 
0.53 $\pm$ 0.14 & 1085& $\pm$ &295 \\ 
094 & 54 & 0.061 & -0.41 $\pm$ 1.38 &  24.88 $\pm$   8.48 & 
0.39 $\pm$ 0.21 &    6& $\pm$ &  3 \\
103 & 32 & 0.200 &  1.31 $\pm$ 0.32 &  34.03 $\pm$   9.81 & 
0.48 $\pm$ 0.14 &  100& $\pm$ & 30 \\ 
117 & 54 & 0.064 & -0.33 $\pm$ 1.86 &  13.00 $\pm$   7.67 & 
0.22 $\pm$ 0.18 &    4& $\pm$ &  3 \\
121 & 27 & 0.310 & -0.13 $\pm$ 0.93 &  27.10 $\pm$   7.06 & 
0.41 $\pm$ 0.15 &  166& $\pm$ & 61 \\
127 & 23 & 0.250 &  0.88 $\pm$ 0.48 &  22.07 $\pm$   6.40 & 
0.31 $\pm$ 0.10 &  103& $\pm$ & 33 \\
131 & 24 & 0.576 &  0.87 $\pm$ 0.54 &  22.87 $\pm$   7.21 & 
0.32 $\pm$ 0.11 &  720& $\pm$ &246 \\
132 & 28 & 0.223 &  1.63 $\pm$ 0.26 &  19.18 $\pm$   5.77 & 
0.25 $\pm$ 0.08 &   74& $\pm$ & 24 \\
134 & 49 & 0.250 &  0.38 $\pm$ 0.83 &  26.33 $\pm$   8.09 & 
0.38 $\pm$ 0.15 &  113& $\pm$ & 43 \\
135 & 32 & 0.520 &  0.93 $\pm$ 0.58 &  17.15 $\pm$   6.46 & 
0.24 $\pm$ 0.10 &  423& $\pm$ &172 \\ \hline
\end{tabular}
\end{minipage}
\end{table*}

\section{The sample}

The UK Deep Survey (Branduardi-Raymont \etal 1994, McHardy \etal 1996) involves
an optical identification programme of a deep ($>$105~ks exposure) ROSAT PSPC
pointed observation. It covers 0.2 deg$^2$ of sky in an area where the Galactic
$N_H$ is low ($\sim$ 6.5 x 10$^{\rm 19}$ cm$^{\rm -2}$) and relatively uniform
(Jones \etal 1995b). The survey reaches a flux limit of 2 x 10$^{\rm -15}$ erg
cm$^{\rm -2}$ s$^{\rm -1}$ (0.5-2 keV). Only sources with offaxis angle up to
15 arcminutes have been used due to the larger positional uncertainty and
possible masking by the detector window support structure at larger radii.
Source searching was carried out in the energy range 0.5 to 2.0 keV (PHA
channels 50 - 200), as the softer band has a larger point spread function
(PSF), higher diffuse Galactic X--ray emission and a larger contribution of
Galactic stars, which complicate the detection of extragalactic sources.

Low resolution (10 - 15 \AA) optical spectra were obtained on the Nordic
Optical Telescope (NOT), the University of Hawaii 88~in Telescope, the Multiple
Object Spectrograph (MOS) on the Canada-France-Hawaii Telescope (CFHT) and the
ISIS spectrograph on the William Herschel (WHT) Telescope (McHardy \etal 1996).

About 90\% of the X--ray sources have been identified and 19 NELGs and 33 AGN
are contained in the sample. The NELGs may consist of a mixture of 
starburst, LINER and Seyfert~2 galaxies, though the precise 
classification within this group remains uncertain in many
cases. 

\section{X--ray spectral analysis}

As most of the UK Deep Survey sources are too faint to warrant construction of
full resolution spectra, we have measured X--ray counts in three bands and
applied a fitting procedure developed by Mittaz \etal (1996) to determine the
spectral parameters of the sources.  The procedure uses a maximum likelihood
technique based on the poissonian distribution of counts (Cash, 1979) to fit a
spectral model together with an accurate background estimate to the observed
source plus background counts.  By using the total number of observed counts
(source plus background), the poissonian nature of the data is correctly
described. This technique is used in preference to the $\chi^{2}$ statistic
which assumes a gaussian probability distribution of the counts; the difference
can become important for low source counts such as those in our sample.  The
spectral fitting procedure has been checked on bright sources taken from the
ROSAT International X--ray Optical Survey (RIXOS). Excellent agreement is found
between the results of the three-band fits to the data and those obtained using
the software package XSPEC applied to data with the full PHA resolution of the
PSPC (Mittaz \etal 1996).

The three spectral bands used in this analysis are defined as: S (PHA channels
8 -- 41 inclusive), H1 (channels 52 -- 90) and H2 (channels 91 -- 201). At the
first stage in the reduction process, we excluded intervals of high master veto
rate and bad aspect solution, and also intervals of anomalously high and low
count rate in the time series of the whole field, leaving 85~ks of `clean'
data.  The source counts are obtained using the software package ASTERIX,
summing all the counts in each band from a circle centred at the source
position. The radius of the extraction circle was nominally 54 arcseconds.
However where the extraction circles of two sources overlap, this radius was
reduced to one half of the distance between the source and its nearest
neighbour. The resulting counts were corrected for the `missing' fraction of
the PSF response using the data of Hasinger et~al. (1994). A large, annular
(4.8 - 10.2 arcminutes radii) region centred on the pointing position of the
field, was used to calculate the average background in each band with the
sources masked out, after correcting for vignetting using the exposure map
provided with the original data.

A two parameter ($\alpha$, Normalization) power-law model with the neutral
hydrogen absorbing column ($N_H$) fixed to the Galactic value  ($6.5 \times
10^{19}$ cm$^{-2}$) was used in the fitting process, leaving one degree of
freedom. The detector response function, the background level and corrections
for vignetting and energy dependent  PSF effects were folded into the fitting
process.

\section {Results and discussion}

Table 1 lists the results of the power-law fits to the data on the 19
individual NELGs in our sample. As noted previously, the absorption was 
fixed at the Galactic value.
The fluxes were obtained from the source counts, using the ROSAT PSPC response
matrix, and correcting for the Galactic $N_H$. We have adopted a value of q$_0$
= 0 and H$_0$ = 50 km sec$^{-1}$ Mpc$^{-1}$ when calculating the X--ray
luminosities. 

The single power-law model with $N_H$ fixed to the Galactic value is an
adequate fit to most of the sources.  However, based on the residuals to the
fit, there is evidence for a somewhat higher absorbing column and a steeper
slope in $\sim$ 10\% of the NELGs and $\sim$25\% of the AGN. There is no
evidence for a systematic trend of NELG spectral slope with, say, luminosity or
redshift (Table~1).

The slope for each individual NELG and AGN in the sample has been plotted
against the total signal in PHA channels 52-201 in Fig 1. The
number distribution of the two source types is shown in Fig.~2. From these
figures it can be seen that the brighter sources are AGN, whilst NELGs only
appear and become important in the faint end of the distribution, below
$\sim$120 counts. Even though there is a significant dispersion of slopes
from source to source, there is an indication even in these data that the
slope of the NELGs is systematically harder than that of the AGN.

\begin{figure}
\leavevmode
\epsfxsize=90mm
\epsfysize=68mm
\epsffile{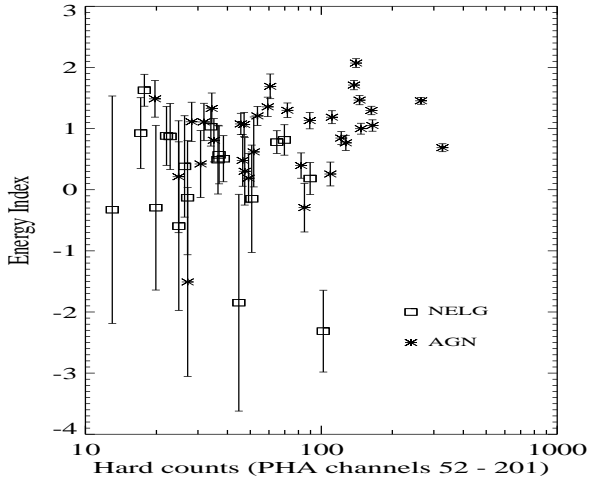}
\caption[]{Energy index versus counts in PHA channels 52 - 201 for AGN and
NELGs in the UK Deep Survey.}
\end{figure}

\begin{figure}
\leavevmode 
\epsfxsize=90mm 
\epsfysize=68mm 
\epsffile{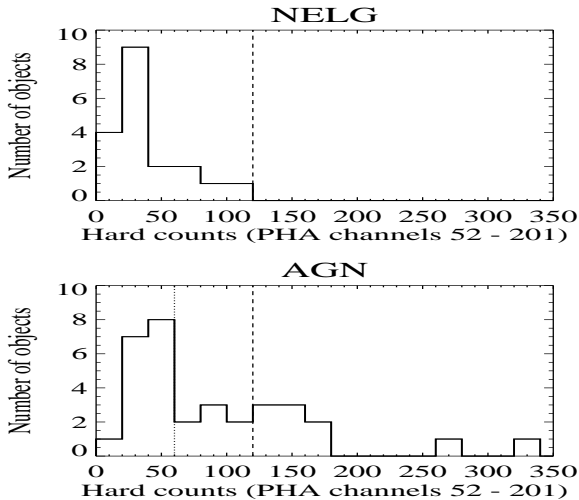}
\caption[]{Number of sources versus counts in PHA channels 52 - 201}
\end{figure}

To investigate further the difference between the X--ray spectra of NELGs and
those of broad line AGN, we co-added the NELG and the AGN counts in each band
and created an average spectrum of these two types of sources, taking into
account the different offaxis angle of each source and correcting for the
vignetting effect. By summing the spectrum of many NELGs and AGN we are
averaging the effects of dispersion in the properties of individual objects,
mimicking the way the unresolved X-ray background is measured. These average
spectra were then fitted in the same way as the individual sources. A
probability distribution for the power-law slope is obtained by projecting 
(integrating) the two dimensional maximum likelihood surface in normalisation
and slope  along the normalisation axis to define probability densities for the
spectral slopes (Cash 1979).

\begin{figure}
\leavevmode
\epsfxsize=90mm
\epsfysize=65mm
\epsffile{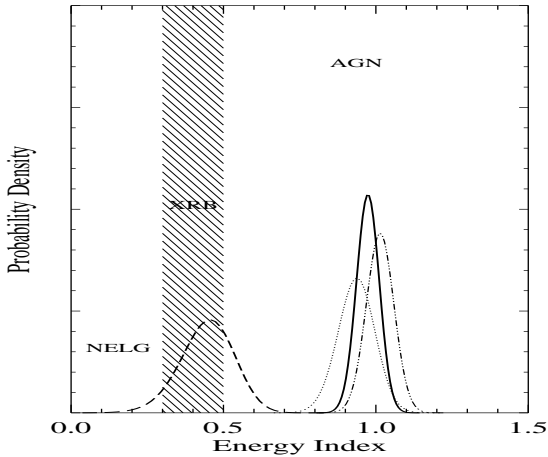}

\caption[]{Probability distribution of AGN and NELG spectral slopes.  The
shaded area represents the X--ray background slope, the width of which reflects
its uncertainty (Gendreau \etal 1995). The dashed line represents the NELGs
distribution. The solid line is used for AGN with less than 120 counts in the
PHA channels 52 - 201. These AGN have been further divided into two subsamples,
the dotted line being used for AGN with less than 60 counts, and the
dash-dotted line representing AGN with counts between 60 and 120. See text for
details.}

\end{figure}

To ensure that there is no bias in the spectrum as a function of count rate
(such as might be introduced by imperfect background subtraction for example)
we have used only AGN with $<$120 counts in order to match the count range of
the NELG sample.  This leaves 23 AGN in the sample. The probability
distribution for the AGN and NELGs energy indices are compared in Fig.~3. We
find that the mean slope of the AGN is $\alpha$=0.96$\pm$0.03 (1-sigma)  while
that of the NELGs is $\alpha=0.45\pm0.09$.  

The error quoted for the mean slope of the AGN and NELG samples is statistical
only, which is appropriate because we are interested in their relative value
and systematic errors would affect both samples in the same way. However,
systematic errors will be relevant when comparing our results with those of
other samples. The dominant cause of possible systematic error in our spectral
extraction procedure is likely to be the background value used. To assess this
we have artificially distorted the background spectral shape and refit the
average source data. For illustration, if we perturb the counts in the softest
and hardest channels by plus and minus 5\% respectively, the slope of both the
mean AGN and NELG samples changes by 0.2. We regard this as an upper limit to
the size of possible systematic effects.

Given that systematic errors in the background will affect the results on faint
sources more than bright sources, we have verified that the AGN and NELG 
populations have different mean slopes by further dividing  the AGN sample into
two groups which have $<$60 counts and counts between 60 and 120 respectively. 
Probability distributions for these two subsamples are also indicated in
Fig.~3. Both AGN subsamples are significantly softer than the NELGs. Moreover
the mean slope found for the AGN is consistent with brighter X--ray AGN samples
selected in the soft X--ray band above $\sim$0.3 keV (Maccacaro \etal 1988,
$\overline{\alpha}$ = 0.95 $\pm$ 0.05; Mittaz \etal 1996, $\overline{\alpha}$ =
1.05 $\pm$ 0.07 ).

We have excluded the ten brightest AGN in our sample from the analysis. If we
include them, the mean AGN slope increases to 1.21 $\pm$ 0.01, increasing the
discrepancy with the NELG results.  However it can be seen from Fig.~1 that
this softening of the mean is primarily caused by 2 of these bright AGN that
have a softer than average spectrum and dominate the counts.

We also note that there are inevitable selection effects in a count-rate
limited sample derived with ROSAT caused by the restricted energy response of
the detector.  We expect
the mean spectrum of source samples to be biased to softer slopes than would be
the case for a flux limited sample selected over a more extended energy range
and with a flatter high energy response.  This should be borne in mind when
comparing our results with the spectrum of the X-ray background measured, say,
with ASCA which should include significant numbers of faint AGN as well as
NELGs.   Nevertheless, since such selection effects would affect both the AGN
and NELG samples, the {\em difference} in the mean spectrum of the two samples
seems secure. 

We also note that direct measurements of the X-ray background with ROSAT may
suggest a somewhat softer slope than determined with ASCA (e.g. 0.7$\pm$0.3,
Branduardi-Raymont et~al.~1994; 0.6$\pm$0.3,  Chen et~al. 1994). However the
uncertainties are relatively large given the restricted energy range of ROSAT,
and there are possible systematic biases due to the uncertain contribution of
Galactic background emission in this band.

\section{Conclusion}

We have integrated the X--ray spectrum of 19 NELGs taken from the UK Deep
Survey and compared this with the spectrum of 23 AGN of similar count rate in
the same sample. We find that the NELG spectrum is harder than that of the AGN
at more than 3~$\sigma$ confidence.

Moreover, the mean spectral slope of the NELGs ($\alpha=0.45 \pm 0.09$) is
consistent with the slope of the X--ray background between 1 and 10 kev
($\alpha = 0.4\pm0.1$; Gendreau \etal 1995) whereas the slope of the AGN at
similar count rates is not ($\alpha \sim 1.0$).

This work is important for understanding the origin of the soft ($< 2$ keV)
X--ray background. Extrapolation of the source number counts suggests that
NELGs are the dominant source population at fluxes four times fainter than the
deep survey flux limit (i.e. at fluxes of $\sim 5\times 10^{-16}$ erg cm$^{-2}$
s$^{-1}$ in the 0.5-2 keV band; Jones \etal 1995a, McHardy \etal 1996).   The
increasing number of these sources at low fluxes and their spectral properties
as shown in this paper, taken together, can reproduce both the flux and
spectrum of the X--ray background.  This result adds considerable weight to the
idea that NELGs are  the major contributor to the  residual unresolved soft
X-ray background.

\vskip 0.4cm
\section{acknowledgements}

We thank the many people who have contributed to the UK deep survey project.
ERC would also like to thank D. Romero and E. Colmenero for their support.
   
\section{References}
\refer Boyle B.J., McMahon R.G., Wilkes B.J., Elvis M., 1995, MNRAS, 272, 462
 315 
\refer Boyle B.J., Shanks T., Georgantopoulos I., Stewart G.C., Griffiths R.E.,
1994, MNRAS, 271, 639
\refer Branduardi-Raymont G., \etal 1994, MNRAS, 270, 947 
\refer Cash W.J., 1979, ApJ, 228, 939 \refer Chen L.W., Fabian A.C. and
Gendreau, K.C. 1995, submitted to MNRAS. 
\refer Fabbiano G.,1989, Ann. Rev. AA, 27, 87 
\refer Fabian A.C. and Barcons, X. 1992, Ann. Rev. AA, 34, 429 
\refer Gendreau K.C., \etal 1995, PASJ, 47, L5 
\refer Georgantopoulos I. \etal 1993, MNRAS, 262, 619 
\refer Hasinger G., \etal 1994, MPE/OGIP Calibration Memo CAL/ROS/93-015 
\refer Hasinger G., Burg R., Giacconi R., Hartner G.,   Schmidt M., Tr\"{u}mper
J. and Zamorani G., 1993, A\&A, 275, 1 
\refer Jones L.R., \etal, 1995a, Proc ``Wide field spectroscopy and the distant
   Universe'', ed Maddox,S. \& Aragon-Salamanca, A., World Scientific Press,
   p346 
\refer Jones, M.H., Rowan-Robinson, M., Branduardi-Raymont, G., Smith,
   P., Pedlar, A., Willacy, K., 1995b, MNRAS, 277, 1587 
\refer Maccacaro T., Goia I.M., Wolter A., Zamorani G. and Stocke J.T., 1988,
ApJ, 326, 680
\refer McHardy I.M., \etal 1996, in preparation 
\refer Mittaz J.P.D., \etal 1996, in preparation 
\refer Romero-Colmenero E., Carrera F.J., Branduardi-Raymont G., Mittaz J.P.D.,
McHardy I.M., Jones L.R. and the~RIXOS~consortium,~1996, Proc. ``R\"{o}ntgenstra- \\hlung from the Universe'', ed.~H.U.Zimmermann,
J.Trumper, \& H.Yorke, MPE report 263, p501.
\refer Shanks T., Georgantopoulos I., Stewart G.C., Pounds K.A., Boyle B.J. and
Griffiths R.E., 1991, Nature 353, 315
\end{document}